# Unexpected crystalline homogeneity from the disordered bond network in La(Cr$_{0.2}$Mn$_{0.2}$Fe$_{0.2}$Co$_{0.2}$Ni$_{0.2}$)O$_3$ films

Matthew Brahlek[1]*, Alessandro R. Mazza[1], Krishna Chaitanya Pitike[1], Elizabeth Skoropata[1], Jason Lapano[1], Gyula Eres[1], Valentino R. Cooper[1], T. Zac Ward[1]

[1]Materials Science and Technology Division, Oak Ridge National Laboratory, Oak Ridge, TN, 37831, USA

Correspondence should be addressed to *brahlekm@ornl.gov

**Abstract**: Designing and understanding functional electronic and magnetic properties in perovskite oxides requires controlling and tuning the underlying crystal lattice. Here we report the structure, including oxygen and cation positions, of a single-crystal, entropy stabilized perovskite oxide film of La(Cr$_{0.2}$Mn$_{0.2}$Fe$_{0.2}$Co$_{0.2}$Ni$_{0.2}$)O$_3$ grown on SrTiO$_3$ (001). The parent materials range from orthorhombic (LaCrO$_3$, LaMnO$_3$ and LaFeO$_3$) to rhombohedral (LaCoO$_3$ and LaNiO$_3$), and first principles calculations indicate that these structural motifs are nearly degenerate in energy and should be highly distorted site-to-site. Despite this extraordinary local configurational disorder on the *B*-site sublattice, we find a structure with unexpected macroscopic crystalline homogeneity with a clear orthorhombic unit cell, whose orientation is demonstrated to be controlled by the strain and crystal structure of the substrate for films grown on (La$_{0.3}$Sr$_{0.7}$)(Al$_{0.65}$Ta$_{0.35}$)O$_3$ (LSAT) and NdGaO$_3$ (110). Furthermore, quantification of the atom positions within the unit cell reveal that the orthorhombic distortions are small, close to LaCrO$_3$, which may be driven by a combination of disorder averaging and the average ionic radii. This is the first step towards understanding the rules for designing new crystal motifs and tuning functional properties through controlled configurational complexity.

The $ABO_3$ perovskite complex oxides exhibit a great diversity of phenomena ranging from strongly correlated electronic phases [1,2], to novel magnetic [3] and multiferroic phases [4,5]. These fundamentally result from the complex interactions among the specific $A$- and $B$-site cations and the relation and interplay with the local bonding geometry. The perovskites are a special materials class because the corner sharing oxygen octahedral units make it extremely manipulatable from a structural perspective [6]. The corner connectivity is critical because it enables the octahedra to have many degrees of freedom, which modifies the $B$-O-$B$ bond angles and bond lengths. This is demonstrated in Fig. 1(a) where an octahedron is rotated about the $x$-axis by angle $\alpha$, $y$-axis by $\beta$ and the $z$-axis by $\gamma$. This then transfers to the surrounding octahedra through the corner connectivity, which can rotate either in-phase or out-of-phase along a particular axis, as indexed in the standard Glazer notation by using a + or – superscript, respectively [6,7]. This so-called pseudocubic motif encompasses many different formal spacegroups [7,8]. Fig. 1(a) gives examples of the orthorhombic $Pbnm$ structure and the rhombohedral $R\bar{3}c$ structure, which are described in Glazer notation $a^-a^-c^+$ and $a^-a^-a^-$, respectively. This structural flexibility enables many routes to manipulate the bond network and ultimately the underlying electronic and magnetic phenomena through epitaxial strain, interfacial control and heterostructuring, or cation substitution.

Isovalent cation substitution has been demonstrated as an effective means of applying internal chemical pressure to perovskites [9–11], but is generally challenging due to the inherent difficulty to stabilize single phase crystals where a single cation sublattice is populated by multiple atomic species, as each species can have a different formation enthalpy [12]. This limitation in synthesis was recently overcome with the discovery of entropy-driven stabilization processes in, so called, high entropy oxides. In these systems, the dominance of the entropy over enthalpy is used to drive the system toward a single homogeneous structure which is randomly populated by constituent cations at the atomic level. Initial work focused on the binary oxides with a rocksalt structure where the cation sublattice was shown to host five or more different elements without quenching disorder [13–15]. This ability to greatly expand configurational complexity in single crystals offers an extremely powerful platform to answer unique questions regarding electronic and magnetic phase formation in the strong disorder limit, while providing new opportunities to create designer crystal lattices using a new set of rules.

Very recently, this approach has been taken in a number of perovskite oxides, where single crystals populated by five or more elements on either the $A$-site or $B$-site sublattice have opened investigations of functional responses in the limit of extreme configurational disorder [16–18]. While it is well known that perovskite functionality is often tightly bound to local structural properties, there are currently no examples in the literature detailing how a multiple populated sublattice might structurally evolve from the local to macro scales. The high configurational disorder in these materials can be expected to lead to complexity in the local bonding environment and is therefore critical to our understanding of the physics that drives functionality. Here we report the structure of the entropy stabilized perovskite system $La(Cr_{0.2}Mn_{0.2}Fe_{0.2}Co_{0.2}Ni_{0.2})O_3$ (L5$BO$) grown on $SrTiO_3$. By considering the average ionic radii as well as utilizing first principles calculations, we show that the large level of configurational disorder should give rise to a phase mixture of rhombohedral and orthorhombic structures. However, we demonstrate that this is not the case. Instead, long range structural order is observed in the form of globally uniform weak distortions in a homogeneous orthorhombic structure; the observed structure is neither identical to a specific parent material nor an average of all parents.

$La(Cr_{0.2}Mn_{0.2}Fe_{0.2}Co_{0.2}Ni_{0.2})O_3$ is used as the entropy stabilized perovskite system in this work. In the parent trenary perovskites, the ionic radii of the cations dictate the crystal structure. In the bulk $LaCrO_3$, $LaMnO_3$, and $LaFeO_3$ are orthorhombic; while $LaCoO_3$ and $LaNiO_3$ are rhombohedral. When considered in series, the change in $B$-site has a strong influence on the fundamental unit cell. Orthorhombic distortion (octahedral tilt angles and $A$-site displacements) evolve from relatively low distortion in $LaCrO_3$, to intermediate in $LaMnO_3$, to highly distorted in $LaFeO_3$, before converting to a rhombohedral structure in $LaCoO_3$ and $LaNiO_3$, as shown in the structural models in Fig. 1(b). For the parent materials, this change in structure is captured by the modification of the tolerance factor $t = (r_A + r_O)/(\sqrt{2}(r_B + r_O))$ [19], which depends solely on the ratio of the constituent ionic radii, with $r_A$, $r_B$ and $r_O$ being the $A$-site, $B$-site, and oxygen radii, respectively. As a function of the *B*-site cation, Fig. 1(c) shows the ionic radii, Fig. 1(d) the tolerance factor, and Fig. 1(e) the pseudocubic lattice parameters. Here, the radii for both the low-spin and high-spin $3^+$ valence states are plotted [20,21], with the open symbols indicating whether the high-spin or low-spin state is taken by the parent material [1]. When the *A*-site is solely populated by La, the specific octahedral rotation pattern and the *B*-O-*B* bond angle and length are driven by the character of the *B*-site cation. As such, the site-to-site disorder on the *B*-site in L5*B*O should give rise to a strong site-to-site variation of the octahedra. This is similarly expected for the La positions, which due to *A*-O covalency distort in the octahedral tunnels [22], as can be seen in Fig. 1(b). The specific choice of *B*-sites of Cr, Mn, Fe, Co, and Ni creates a situation which spans weak to highly disordered orthorhombic and rhombohedral materials. Thus, there is an expectation that the local bonding network in La(Cr$_{0.2}$Mn$_{0.2}$Fe$_{0.2}$Co$_{0.2}$Ni$_{0.2}$)O$_3$ should contain a large distribution of bond angles and lengths that would give rise to a highly disordered unit cell that could range from cubic, to rhombohedral or orthorhombic.

To make the argument regarding the energetically favored structure quantitative, we have performed density functional theory (DFT) calculations on a 2×2×10 supercell where the atomic sites were chosen quasi-randomly, as shown in Fig. S1; further details can be found in Ref. [23]. Here we examined the relative stability and structure of the cubic, orthorhombic and rhombohedral phases after the structures were allowed to relax. From the first principles calculations we find that the lowest energy state is rhombohedral with the orthorhombic unit cell being slightly higher at 3 meV per formula unit (FU) and the cubic phase is significantly larger at about 149 meV/FU. This implies that, within the error in the calculation, the rhombohedral and orthorhombic structures are degenerate. The higher energy of the cubic phase is likely a reflection of the atomic size of the La relative to the average of the *B*-sites, as captured by the tolerance factor.

Based on the cubic, rhombohedral, and orthorhombic unit cells we analyzed the *B*-O-*B* disorder in bond angles of each supercell (Fig. 2). Here, we see that in the rhombohedral and orthorhombic unit cells, the distribution of bond angles centered around 161° with peak widths of ±2°, while the cubic phase is peaked at 180° with a tail extending below this by about 3-4° (see Fig. S2 for cation positions in Ref. [23]). In all cases, the broad distribution suggests a disordered ground state, which should appear in a diffraction experiment as a strong peak-broadening. Nevertheless, the DFT calculations indicate the possibility of either rhombohedral or orthorhombic phases; regarding why the experimental structure is found to be orthorhombic is discussed at the end of the paper.

Structural refinement was performed on a 47 nm L5*B*O film grown using pulsed laser epitaxy (PLE) on a 5×5 mm$^2$ SrTiO$_3$ (001) substrate using X-ray diffraction (XRD); further experimental details can be found in Ref. [23]. Figure 3 shows a $2\theta$-$\omega$ scan about the 002 Bragg reflection of the film and the substrate. The sharp peak at $2\theta \approx 46.5°$ is the SrTiO$_3$ substrate peak corresponding to a *c*-axis lattice parameter of 3.905 Å. The broader peak on the right side at about $2\theta \approx 47.1°$ is the 002 reflection of the L5*B*O film, which corresponds to an out-of-plane lattice parameter of 3.853 Å. These results are consistent with previous results [17] and are reproduced across many samples. Furthermore, 20+ Laue oscillations about the 002 reflection are visible, which indicate that the film is atomically flat and uniform in thickness. In Fig. 3(b) a 2-dimensional reciprocal space map about the 113 reflection is shown, where the vertical axis is the out-of-plane reciprocal lattice vector, $q_\perp$, and the horizontal is the in-plane reciprocal lattice vector, $q_\parallel$. This enables comparison between the in-plane and out-of-plane spacing of both the L5*B*O film and SrTiO$_3$. From this data, it is clearly seen that the film and the substrate share a common in-plane lattice parameter of 3.905 Å—confirmation that the film is coherently strained. Together this data shows that the unit cell is highly uniform despite the high cation configurational disorder inherent to these materials.

With the data shown in Fig. 3, the Poisson ratio can be used to infer the bulk lattice parameter of L5*B*O. This is significant since there are no bulk single crystals; PLE thus far offers the only route to synthesize single crystals of these materials. When a material with lattice parameter $a_f$ is strained to an in-plane lattice parameter of $a_\parallel$ (= 3.905 Å), the out of plane lattice parameter is given by

$$a_\perp = \left(1 - \left(\frac{2\nu}{1-\nu}\right)\left(\frac{a_\parallel - a_f}{a_f}\right)\right) a_f. \qquad (1)$$

In this equation $a_\perp$, $a_\parallel$ are determined experimentally, $\nu$ is Poisson's ratio, typically between 0.2-0.4 for perovskites. This can be readily solved for the unstrained lattice parameter of the film

$$a_f = \frac{a_\perp + \nu(2a_\parallel - a_\perp)}{1+\nu}. \qquad (2)$$

Substituting the experimental values for $a_\parallel$ and $a_\perp$ and taking $\nu$ to be 0.3, yields an unstrained lattice parameter of 3.877 Å (for $\nu = 0.2$, $a_f \approx 3.870$, and $\nu = 0.4$, $a_f \approx 3.883$). To contextualize this value, we can compare it to the relative lattice parameters for the parent components: the pseudocubic lattice parameters (taken by calculating the cube-root of the volume per pseudocubic unit) are $LaCrO_3$ is 3.89 Å [24], $LaMnO_3$ is 3.93 Å [25], $LaFeO_3$ is 3.93 Å [26], $LaCoO_3$ is 3.82 Å [27], $LaNiO_3$ is 3.84 Å [26], which are plotted in Fig. 1(e). These yield an average value of 3.88 Å, which agree well with the experimental value of 3.877 Å extracted from the data in Fig. 3(a-b). This implies that the lattice parameter is determined by the volume packing of the perovskites. Next, we move on to characterize the substructure of the pseudocubic unit cell.

The systematic extinctions of specific reflections enable characterization of unit cell symmetry. For pseudocubic perovskite cells, these are captured by half-order reflections where Miller indices $H$, $K$, and $L$ are mixtures of half-integers and integers. For the possible unit cells, L5$B$O is likely to take one of three forms: (1) A cubic cell which would admit no half-order reflections. (2) A rhombohedral cell which would exhibit peaks such as $\{h/2, h/2, k/2\}$, where $h$ and $k$, herein, are odd integers and $h \neq k$, and peaks such as $\{h/2, h/2, h/2\}$ and $\{h/2, k/2, l\}$ are systematically extinct. (3) An orthorhombic cell which would exhibit peaks such $\{h/2, h/2, k/2\}$, $\{h/2, h/2, h/2\}$ and $\{h/2, k/2, l\}$. These guidelines, summarized in detail in Ref. [6,7,28–30], enable unique characterization of the unit cell. As shown in Fig. 4(b), for example, the $\{h/2, k/2, l\}$ family of peaks are observed to be non-zero, which clearly indicate that L5$B$O is orthorhombic with space group $Pbnm$. Moreover, for materials within this space group, there are three ways the film can orient on a substrate, as shown schematically in Fig. 4(b). In standard Glazer notation, these orientations are referred to as $a^-a^-c^+$, $a^-a^+c^-$, and $a^+a^-c^-$. These orientations can be distinguished experimentally by measuring the $\{h/2, k/2, l\}$ family of reflections, which are zero unless the + axis is along the direction of the integer index. For example, for a film with an orientation of $a^-a^-c^+$, the 1/2 3/2 1 peak is non-zero whereas the 1/2 1 3/2 and the 1 1/2 3/2 will both be zero. As shown in Fig. 4(a) and schematically illustrated in Fig. 4(d), in tensile strain on $SrTiO_3$ L5$B$O clearly exhibits a $a^-a^-c^+$ pattern since the 1/2 3/2 1 is non-zero, while the 1/2 1 3/2 and the 1 1/2 3/2 are absent. In contrast, under compressive strain on $(La_{0.3}Sr_{0.7})(Al_{0.65}Ta_{0.35})O_3$ (LSAT), 1/2 3/2 1 is zero whereas 1/2 1 3/2 and 1 1/2 3/2 are both non-zero, as shown in Fig. 4(b,e), which indicates that the orthorhombic unit cell has been rotated in-plane by 90° and is mixture of $a^-a^+c^-$ and $a^+a^-c^-$. Further, under compressive strain on orthorhombic $NdGaO_3$ (110) the 1/2 3/2 1 and 1/2 1 3/2 reflections are zero, whereas the 1 1/2 3/2 is non-zero. This indicates that the film is a single-domain $a^+a^-c^-$ as shown in Fig. 4(c,f).

The specific orthorhombic orientation ($a^-a^-c^+$, $a^-a^+c^-$, or $a^+a^-c^-$) a perovskite film takes is determined by a combination of minimizing the interface energy associated with the underlying orientation imprinted by the substrate and minimizing the overall strain energy [31]. For example, if the substrate has a tilt pattern --+ (for example, orthorhombic $R$GaO$_3$ or $R$ScO$_3$ with surface perpendicular to the 001 orthorhombic direction, where $R$ is in the rare earth series) lowering interfacial energy will tend to favor a --+ orientation of the film since it will be a lower net energy compared to the additional cost to rearrange the local bond network. However, a situation can arise when the film has an orthorhombic unit cell with a large anisotropy in the lattice parameters. Then, it can be net lower energy if the film changes rotation pattern to minimize the strain energy. For example, consider a material with a large anisotropy among the pseudocubic lattice parameters, i.e. for the pseudocubic $a,b > c$ (this comparison is best made in the pseudocubic bases since the orthorhombic $Pbnm$ $c$-axis is two pseudocubic cells whereas the $a,b$ are approximately $\sqrt{2}$ unit cells); under tensile strain on a substrate with a +-- pattern, interfacial templating will tend to favor +--. However, minimizing strain energy will prefer the --+ orientation. This is, for example, observed in $LaVO_3$ grown on $DyScO_3$ (110) [32]. For the present situation, the $SrTiO_3$ and LSAT substrates are cubic and exhibits an $a^0a^0a^0$ cubic phase, therefore there is no templating, hence, strain and anisotropy of the bulk unit cell determine the $a^-a^-c^+$ orientation on $SrTiO_3$ and $a^-a^+c^-$ and $a^+a^-c^-$ on LSAT.

This is important because the bulk unit cell for L5$B$O is unknown. The fact that under tensile strained L5$B$O forms the $a^-a^-c^+$ structure on SrTiO$_3$ and under compressive strain form $a^-a^-c^-$ and $a^+a^-c^-$ on LSAT shows that the unit cell of the parent material has anisotropy in the orthorhombic lattice parameters. Specifically, the pseudocubic $c$ axis is shorter than $a$ and $b$. Finally, at very similar strain state as LSAT, orthorhombic NdGaO$_3$ (110) is found to stabilize the single domain $a^+a^-c^-$, which can be directly seen as both the film and substrate 1 1/2 3/2 reflections are found in close proximity in Fig. 4(c). Overall, this shows that the substrate parameters (strain, and structure) enable full control of the orientation and domain structure of the unit cell of L5$B$O.

Herein we focus only on L5$B$O film grown on SrTiO$_3$. The $a^-a^-c^+$ orientation is further characterized by displacements of the La atoms away from their high-symmetry positions, which is associated with the octahedral rotation patterns. This is shown schematically in Fig. 5(a), where the displacements can occur in one of four ways, labeled 1R, 2R, 1L, and 2L which are the allowed combinations of displacements of the La along the <110> and <1-10> pseudocubic directions, $d_1$, and the <100> and <010> pseudocubic directions, $d_2$ (see Ref. [29] for more details). All four of these displacement patterns may be equally favored depending on the effects of templating, strain, etc., and segregate into domains. This can be experimentally checked by measuring the ±1/2 ±1/2 3/2 reflections and comparing the relative intensities. This family of peaks is sensitive to the cooperative displacement among both the $A$-site cations and the oxygen octahedra, and definitely probes the relative domain fraction [29]. Equal intensity within the family implies equal domain populations, $D_j$. As shown in Fig. 5(b), the four ±1/2 ±1/2 3/2 reflections are indeed of equal intensity, which shows equal populations of the 1R, 2R, 1L and 2L domains. This is likely determined by a lack of rotation-pattern-templating by the SrTiO$_3$ substrate.

Given the basic information concerning the unit cell, we can move on to characterize the orthorhombic substructure. In contrast to bulk materials that utilize methods such as Rietveld refinement to locate the atom positions within the unit cell by matching calculated intensities to experimental values, thin film perovskites are inherently difficult due to the small sample volume combined with very weak intensity of the half-order reflections. As has been shown previously [29,30], this can be done by creating a pseudocubic unit cell and parameterizing the oxygen positions with the octahedral rotation angles $α$, $β$, and $γ$, shown in Fig. 1(a), and $d_1$ and $d_2$ for the $A$-site displacements, shown in Fig. 5(a). This enables the structure factor to be calculated for all reflections, which, in turn, can be used to fit experimental data. To discuss the parameterized structure factor we follow closely Ref. [29], with several key modifications. The full structure factor for a perovskite unit cell is given by

$$F_{HKL} = M_O f_{O^{2-}} \sum_{n=1}^{24} e^{2\pi i(Hu_n + Kv_n + Lw_n)} + BM_A f_A \sum_{m=1}^{8} e^{2\pi i(Hu_m + Kv_m + Lw_m)} + M_B f_B \sum_{p=1}^{8} e^{2\pi i(Hu_p + Kv_p + Lw_p)},$$ (3)

where $f_{O^{2-}}$, $f_A$, and $f_B$ are the atomic form factors for oxygen and the $A$-site and $B$-site cations, respectively. In the exponentials, $u_j$, $v_j$, and $w_j$ are the atomic positions within the unit cell which are parameterized by $α$, $β$, $γ$, $d_1$ and $d_2$ (for explicit functional forms see tables in Ref. [29]). It is standard to take $α = β$ since the in-plane lattice parameters are strained to SrTiO$_3$, as shown in the reciprocal space map in Fig. 3(b), which leads to equal bond angles along the [100] and [010] [29]. The functional forms for the atomic form factors are taken from Ref. [33] for oxygen in the 2$^-$ state and Ref. [34] for the cations in the 3$^+$ state. For the $B$-site cations, the form factors are very similar in overall magnitude, and $f_B$ is taken as the average of $Cr^{3+}$, $Mn^{3+}$, $Fe^{3+}$, $Co^{3+}$ and $Ni^{3+}$ for simplicity. Finally, $M_i$ is the Debye-Waller factor, which has not yet been included in the literature, yet we have found it is necessary to include this term to fit peaks at higher $q$ (see Ref. [23] for specific details as well as Ref. [35]). From this, the peak intensity for a given reflection, $I$, can be calculated using the square magnitude of the structure factor

$$I = I_0 B_F L_p \sum_{j=1}^{4} D_j F_{HKL} F_{HKL}^*,$$ (4)

Where $I_0$ is the normalization intensity taken to be the 103 peak, $B_F$ is the beam footprint correction, which was calculated numerically (see Ref. [23]), $L_p = 1/sin(2θ)$ is the Lorentz polarization correction, and $D_j$ are the previously mentioned domain populations (here $D_{1R,1L,2R,2L} = 25\%$, as found in Fig. 5(b)). The specific reflections used, shown on the horizontal axis of Fig. 6(a), were chosen to have diverse dependence on the structural parameters. An iterative fitting procedure was used to minimize the total squared error between the experimental data and the structure factor. Specifically, an initial set of $\alpha$, $\gamma$, $d_1$, and $d_2$ were used to calculate the square-error, which was then minimized with respect to $d_2$ (using the individual peaks highlighted at the top of Fig. 6(a)). This value of $d_2$ was used to calculate a square-error which was subsequently minimized with respect to $d_1$ (again using the peaks highlighted at the top of Fig. 6(a)). This was then repeated for $\gamma$ and then $\alpha$, which completed a single loop. The full loop was then repeated $n$-times until the values of $\alpha$, $\gamma$, $d_1$ and $d_2$ converged, which was typically 3-4 iterations, but here we used $n = 15$. This procedure was tested by calculating intensities for sets of $\alpha$, $\gamma$, $d_1$ and $d_2$ which were subsequently fit using this algorithm and found to be very effective at reproducing the initial set of $\alpha$, $\gamma$, $d_1$ and $d_2$; this is detailed in Ref. [23] in Fig. S4.

The experimental intensities are shown in Fig. 6(a) as gray bars, and the $H K L$ indices are listed at the bottom. This data shows that the intensities range across several orders of magnitude, which highlights the challenge of this measurement. Figure 6(b-e) shows the results of the iterative fitting procedure as a function of $n$, the number of iterations through the fitting loop. Figure 6(b-c) show the results of $\alpha$ versus $\gamma$, Fig. 6(b), and individually $\alpha$ versus $n$ and $\gamma$ versus $n$, Fig. 6(c). Similar data is shown for $d_1$ and $d_2$ in Fig. 6(d-e). Here multiple starting parameters were used. It is seen that regardless of the initial parameters the algorithm quickly finds the best fit, which yields $\alpha \approx 6.3°\pm0.5°$, $\gamma \approx 4.3°\pm0.5°$, $d_1 \approx 0.9$ pm$\pm0.1$ pm and $d_2 \approx 2.4$ pm $\pm0.1$ pm. The error is estimated based on a combination of the estimated error in the measured intensities and simulations [29]. The fit intensities are plotted next to the experimental data in Fig. 6(a) as red bars, where the fit matches very well to the experimental data. Note that there are several peaks shown in Fig. 6(a) that are not included in the fitting procedure, and the calculated intensities for these peaks match very closely to the experimental values, which further indicates the goodness of fit and reliability of the final values.

To understand the implications of the current results we need to draw a comparison between the parent materials and the final structure, as shown in Fig. 7 projected along the pseudocubic $a$, $b$ and $c$ axes. The current results show that the orthorhombic distortions are weak and the structure is highly homogeneous. The key question raised by this data is what is the competition or correlation among average atomic radii and configurational disorder in determining the unit cell symmetry and level of distortion? Considering this in the context of the parent structures, ionic radii and the tolerance factor provides insight:

(1) Importance of the ionic radii: As shown in Fig. 1(e), the lattice parameters of the L5$B$O unit cell are close to the expectation of the average ionic radii based on those of the bulk (the red dashed line $R \approx 0.6$ Å is based on the red open circles Fig. 1(c)). Moreover, the measured orthorhombic distortions ($B$-O-$B$ bond angles of 163° along the $x$-$y$ axes and 161° in along the $z$) is slightly more than the average of the of the parents (161° along in the $x$-$y$ plane and 161° along the $z$), and agree with the first principles calculations. These observations are consistent with the average structure being determined by the average ionic radii, and the reduced distortion being the result of disorder averaging. However, there are some subtle but important differences in the structure relative to the average of the parent materials that must be further considered.

(2) Origin of orthorhombic structure: Assuming the structure is driven by the average ionic radii, the tolerance factor would fall very close to the values for LaCrO$_3$, LaMnO$_3$, LaCoO$_3$, and LaNiO$_3$ and place L5$B$O close to the phase boundary between orthorhombic and rhombohedral. This is further corroborated by the DFT calculations that predicts both structures are nearly degenerate in energy. As such, it is surprising that the material clearly adapts a macroscopically pristine orthorhombic structure. There are several mechanisms that could drive this behavior. Very large local distortions centered on the Mn and Fe octahedra may be strong enough to drive macroscopic stabilization of one structural motif. Furthermore, since these are grown using PLE on SrTiO$_3$ (001) either the specific strain, the kinetics within the PLE plume or on the growing surface, or the growth temperature and cooling rate may all combine to favor the orthorhombic structure. A detailed growth study will shed further light on this.

(3) Local valence state rearrangement: The transition metal cations in these materials are capable of taking several different charge states with a 6-fold coordination as well as both high-spin and low-spin configurations; this may enable nearest neighbor chemical pressure to rearrange local charge states to dissipate regions of high distortion thereby leading to collective-driven structural uniformity. For example, the high-spin and low-spin states in $Co^{3+}$ are nearly degenerate in energy and may fluctuate based on its local environment; Mn readily takes a $3^+$ or $4^+$ valence state with both high-spin and low-spin configuration which could be balanced by, for example, Ni which tends to readily favor a $2^+$ state. This scenario is very likely as valence state rearrangement readily occurs in double perovskites $La_2CoMnO_6$ and $La_2NiMnO_6$ [36–39]. These observations highlight the need for future experimental studies, such as extended X-ray absorption fine structure (EXAFS) and X-ray absorption spectroscopy (XAS), that could provide details on the local bonding environment and charge states, which should guide theoretical studies to better understand the physics of structural formation in configurationally complex systems.

To close, we have refined the structure of single crystal, entropy stabilized perovskite oxide films of $La(Cr_{0.2}Mn_{0.2}Fe_{0.2}Co_{0.2}Ni_{0.2})O_3$ grown on $SrTiO_3$. Individually, the parent materials take a variety of structures—from weakly distorted orthorhombic ($LaCrO_3$) to the highly distorted orthorhombic ($LaMnO_3$ and $LaFeO_3$) to rhombohedral ($LaCoO_3$ and $LaNiO_3$). Based on the average it is expected that L5*B*O is close to the phase boundary between orthorhombic and rhombohedral. It is found that the material is highly homogeneous and adapts an orthorhombic structure that is only weakly disordered; the observed structure is neither identical to a specific parent material nor an average of all parents. This leaves many experimental and theoretical questions open regarding how configurational disorder determines the structure. The measurements clearly show that the distributions of ionic radii are important, but exactly how the individual chemistries of the *B*-sites combine and interact on the highly disordered local scale to stabilize the global structure cannot be understood with simple averaging models and will require future spectroscopic and further first principles investigations to unravel. However, addressing these challenges will enable designing materials with very specific bond-angles, lattice parameters, and, thus, local electronic and magnetic environments by selection of a particular set of *A*- or *B*-site cations. For example, this will then enable (1) understanding how to use disorder to drive new emergent electronic and magnetic phenomena in correlated perovskites, (2) enable actively designing and controlling such states by manipulating the local bonding environment, and (3) these can be used as passive elements at epitaxial heterointerfaces where octahedral rotation patterns, customized using high configurational disorder, can be templated into a second material to drive new functionality. This exciting new class of materials highlights many new directions to design and tune functional properties in new ways to answer fundamental questions in the limit of high configurational disorder.

## Acknowledgements

This work was supported by the U.S. Department of Energy (DOE), Office of Science, Basic Energy Sciences, Materials Sciences and Engineering Division. We thank Jiaqiang Yan for insightful discussions.

**Figure 1**

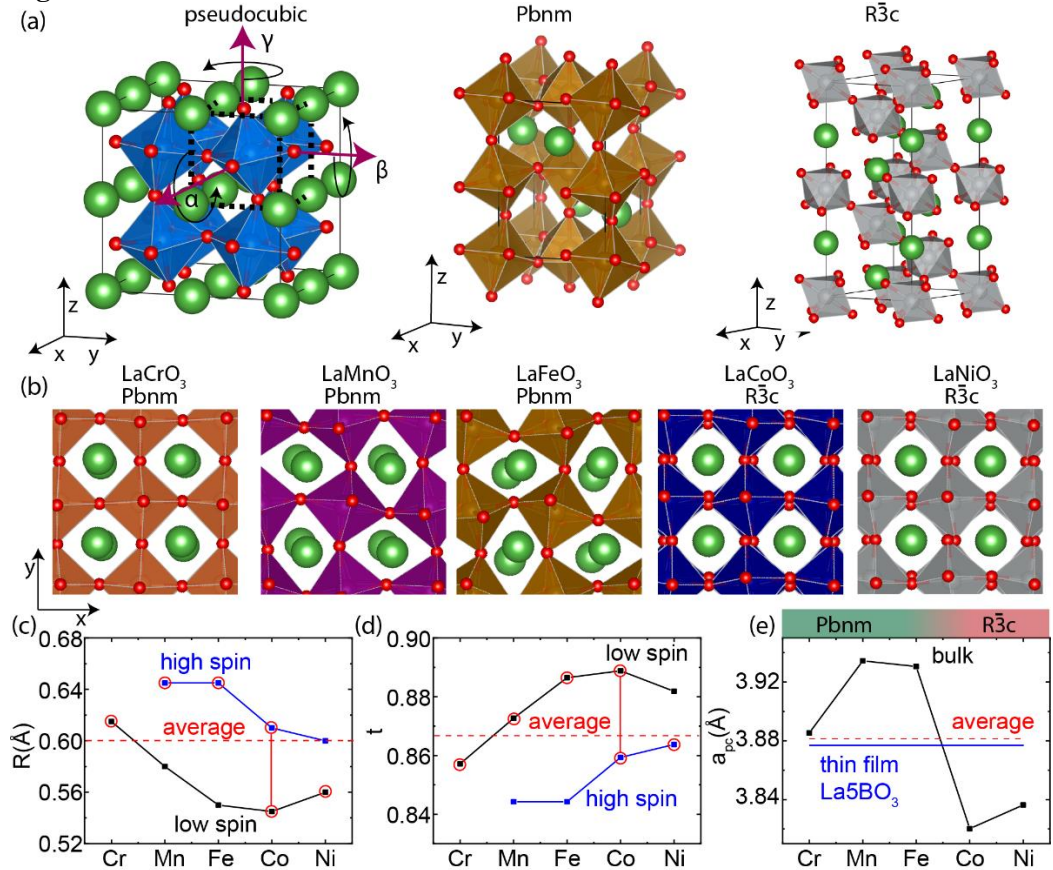

**Fig. 1**. (a) Typical unit cells for perovskites. Left: A pseudocubic cell where the dashed lines highlight a single oxygen octahedra which can be rotated about the *x*-axis by angle $\alpha$, *y*-axis by $\beta$ and the *z*-axis by $\gamma$. Center and right: The formal orthorhombic unit cells for the *Pbnm* and $R\bar{3}c$ structures, respectively. (b) Projections of the unit cell down the <001> pseudocubic direction for the parent compounds of La(Cr$_{0.2}$Mn$_{0.2}$Fe$_{0.2}$Co$_{0.2}$Ni$_{0.2}$)O$_3$. (c-e) Ionic radii (c), tolerance factor (d) and pseudocubic lattice parameter (e) versus *B*-site cation. The solid symbols are data taken from Refs. [20,21], the open symbols represent experimental values of the parent structures from Ref. [1], and the lattice parameters from Refs. [24–27], and the vertical line on Co indicates that both high spin and low spin may be favored.

**Figure 2**

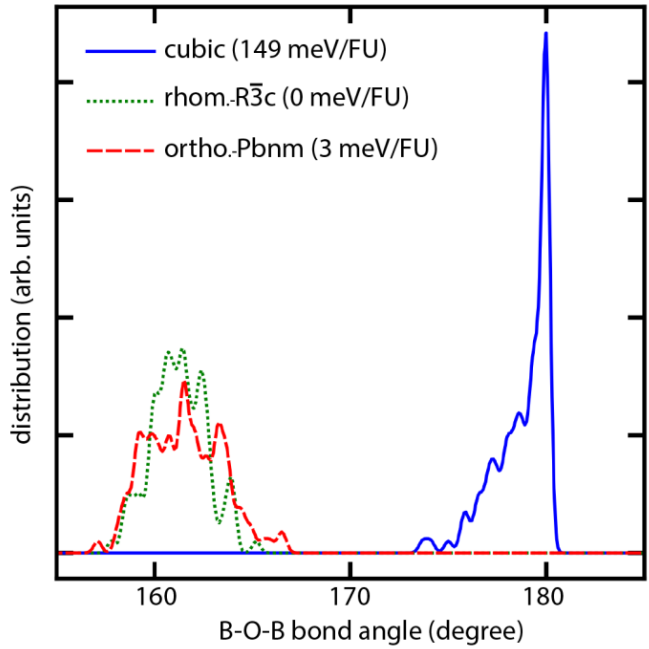

**Fig. 2.** Distribution of the *B*-O-*B* bond angles across the supercell calculated using density functional theory for L5*B*O. This shows three structural motifs where the solid blue line is cubic, the dotted green line is rhombohedral, and the dashed red line is orthorhombic. The number listed in the parenthesis is energy for the different structural motifs relative to the lowest, which is the rhombohedral phase.

**Figure 3**

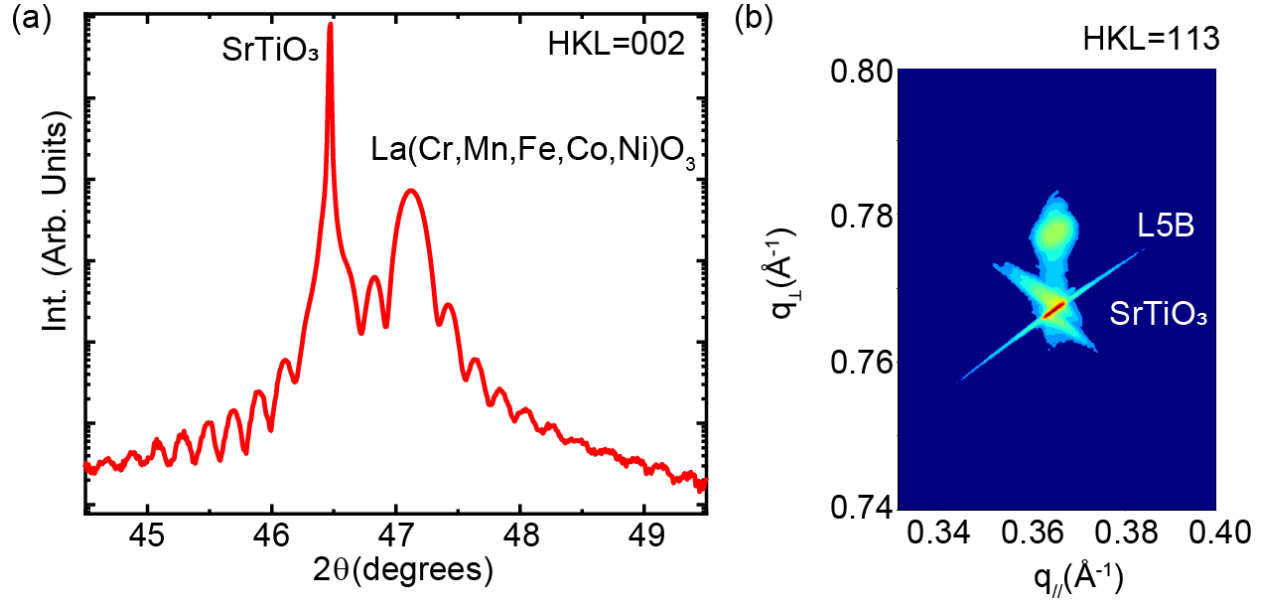

**Fig. 3.** (a) X-ray $2\theta$-$\omega$ scan about the 002 peak for La(Cr$_{0.2}$Mn$_{0.2}$Fe$_{0.2}$Co$_{0.2}$Ni$_{0.2}$)O$_3$ on SrTiO$_3$. (b) Reciprocal space map about the 113 peaks of La(Cr$_{0.2}$Mn$_{0.2}$Fe$_{0.2}$Co$_{0.2}$Ni$_{0.2}$)O$_3$ film and SrTiO$_3$.

**Figure 4**

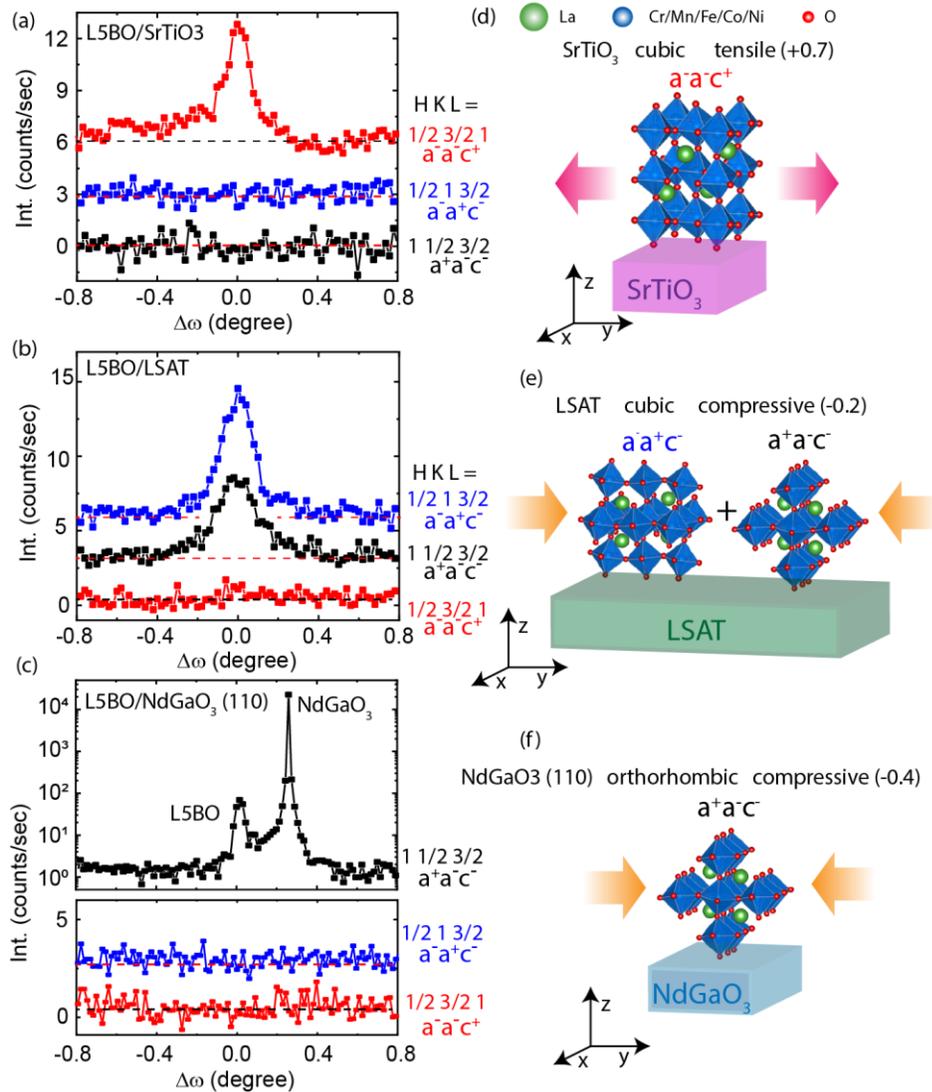

**Fig. 4.** (a-c) X-ray scans of the {1/2 3/2 1} family of reflections corresponding to the possible $a^-a^-c^+$, $a^-a^+c^-$ and $a^+a^-c^-$ orientations for L5$B$O grown on SrTiO$_3$ (a), LSAT (b) and NdGaO$_3$ (110) (c). (d-f) Schematics showing each substrate and the orientation of the orthorhombic unit cell of L5$B$O on SrTiO$_3$ (d), LSAT (e) and NdGaO$_3$ (f). The labels indicate the substrate, its crystal structure, and the strain state relative to the bulk lattice parameters of L5$B$O mentioned in the text, schematically represented as inward (compressive) and outward (tensile) pointing arrows.

**Figure 5**

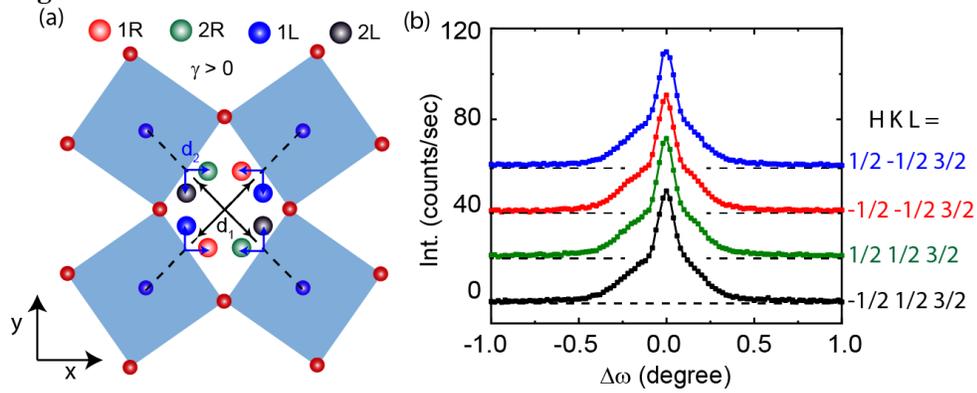

**Fig. 5** (a) Pattern and possible La-displacement patterns labeled 1R, 2R, 1L and 2L. (b) X-ray scans of the {1/2 1/2 3/2} family for L5$B$O grown on SrTiO$_3$ which are used to distinguish the domain patterns in (a).

**Figure 6**

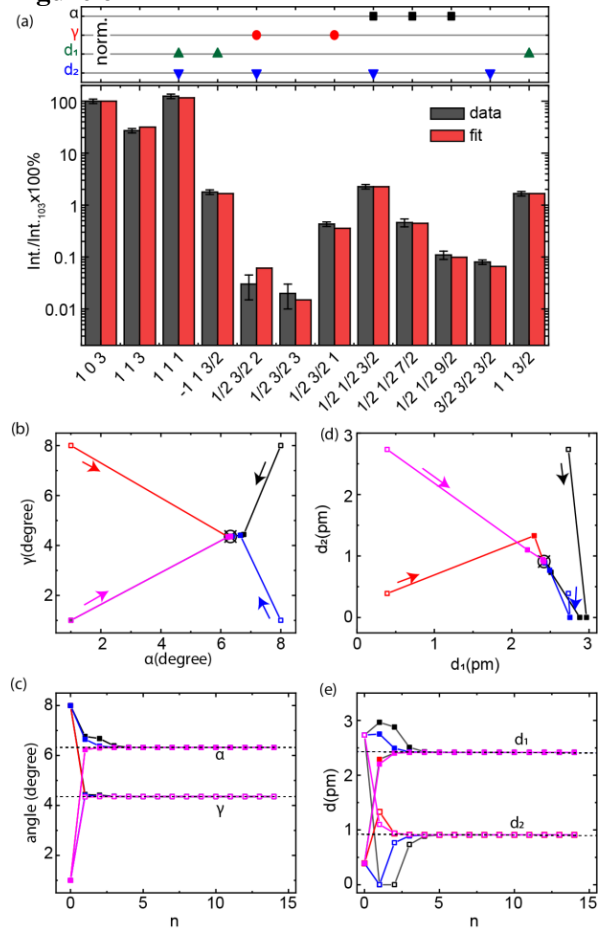

**Fig. 6.** (a) Summary of experimental and fit intensities for L5$B$O on SrTiO$_3$. Bottom axis shows the *H K L* indices for all reflections measured. Top: A symbol indicates that a given reflection was used to fit a given parameter (vertical axis). Middle: Bar chart showing experimental intensity (gray) and the final fit values (red). (b-c) $\gamma$ versus $\alpha$ (c) and $\gamma$ and $\alpha$ versus *n*, the iteration number during the fitting algorithm. (d-e) $d_1$ versus $d_2$ (d) and $d_1$ and $d_2$ versus *n* (e). In (b,d) the open squares are the starting values for the fit algorithm and the crossed-circles are the final values, which are shown in (c,e) as dashed lines.

**Figure 7**

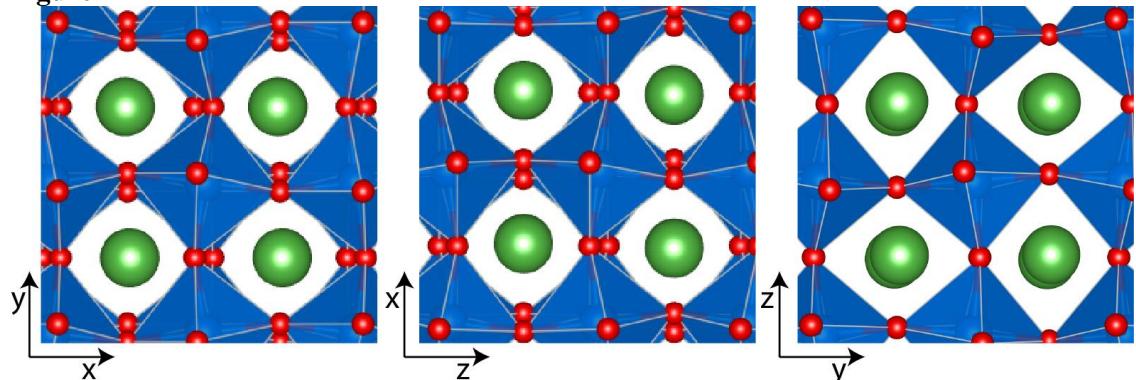

**Fig. 7.** (a-b). The final orthorhombic structure for La(Cr$_{0.2}$Mn$_{0.2}$Fe$_{0.2}$Co$_{0.2}$Ni$_{0.2}$)O$_3$ L5$B$O on SrTiO$_3$ projected along the pseudocubic axes: <100> film direction (left), <010> (middle) and <001> (right).

# Supplemental materials

Unexpected crystalline homogeneity from the disordered bond network in La(Cr$_{0.2}$Mn$_{0.2}$Fe$_{0.2}$Co$_{0.2}$Ni$_{0.2}$)O$_3$ films

Matthew Brahlek[1]*, Alessandro R. Mazza[1], Krishna Chaitanya Pitike[1], Elizabeth Skoropata[1], Jason Lapano[1], Gyula Eres[1], Valentino R. Cooper[1], T. Zac Ward[1]

[1]Materials Science and Technology Division, Oak Ridge National Laboratory, Oak Ridge, TN, 37831, USA

Correspondence should be addressed to *brahlekm@ornl.gov

**Methods: Density Functional Theory**

First-principles calculations were performed using density functional theory (DFT) as implemented in the Vienna Ab initio Simulation Package (VASP) [1,2] with the Perdew-Burke-Ernzerhof for solids (PBEsol) [3] for exchange and correlation. The projector-augmented plane-wave method [4,5] with a 600 eV energy cutoff was used. For a pseudocubic unit cell containing one formula unit of La(Cr$_{0.2}$Mn$_{0.2}$Fe$_{0.2}$Co$_{0.2}$Ni$_{0.2}$)O$_3$ (L5$B$O), a zone-centered $8 \times 8 \times 8$ Monkhorst-Pack $k$-point mesh was used for the Brillouin zone (BZ) integrations. A large supercell was required to model the possible structures for L5$B$O. Therefore a $2 \times 2 \times 10$ supercell, containing 40 L5$B$O formula units (FU), was used and the $k$-point mesh was accordingly scaled down to $4 \times 4 \times 1$. The five cations were quasi-randomly assigned to the particular $B-$site using the special quasi-random structures (SQS) algorithm introduced by Zunger et. al. [6] and implemented in the Alloy Theoretic Automated Toolkit (ATAT) open source software [7]. Figure S1 shows the SQS models for (a) rhombohedral, (b) orthorhombic and (c) cubic structures. In all DFT calculations, internal ionic positions were relaxed until the forces were less than 0.005 eV/Å. The lattice constants were relaxed until the appropriate stress tensor components were less than 1 kbar. To account for the Coulomb interactions within the partially filled $d$–orbitals of the Cr, Mn, Fe, Co and Ni cations, the DFT+$U$ using the simplified (rotationally invariant) approach was used [8]. The value of $U$ used for the $B-$site cations in parent perovskites range between $U = 2.9$ to 7.0 [9–11]. $U = 4$ eV was used for all the $B-$site cation $d$–states. $G-$type antiferromagnetic ordering was initiated on the $B-$site sublattice.

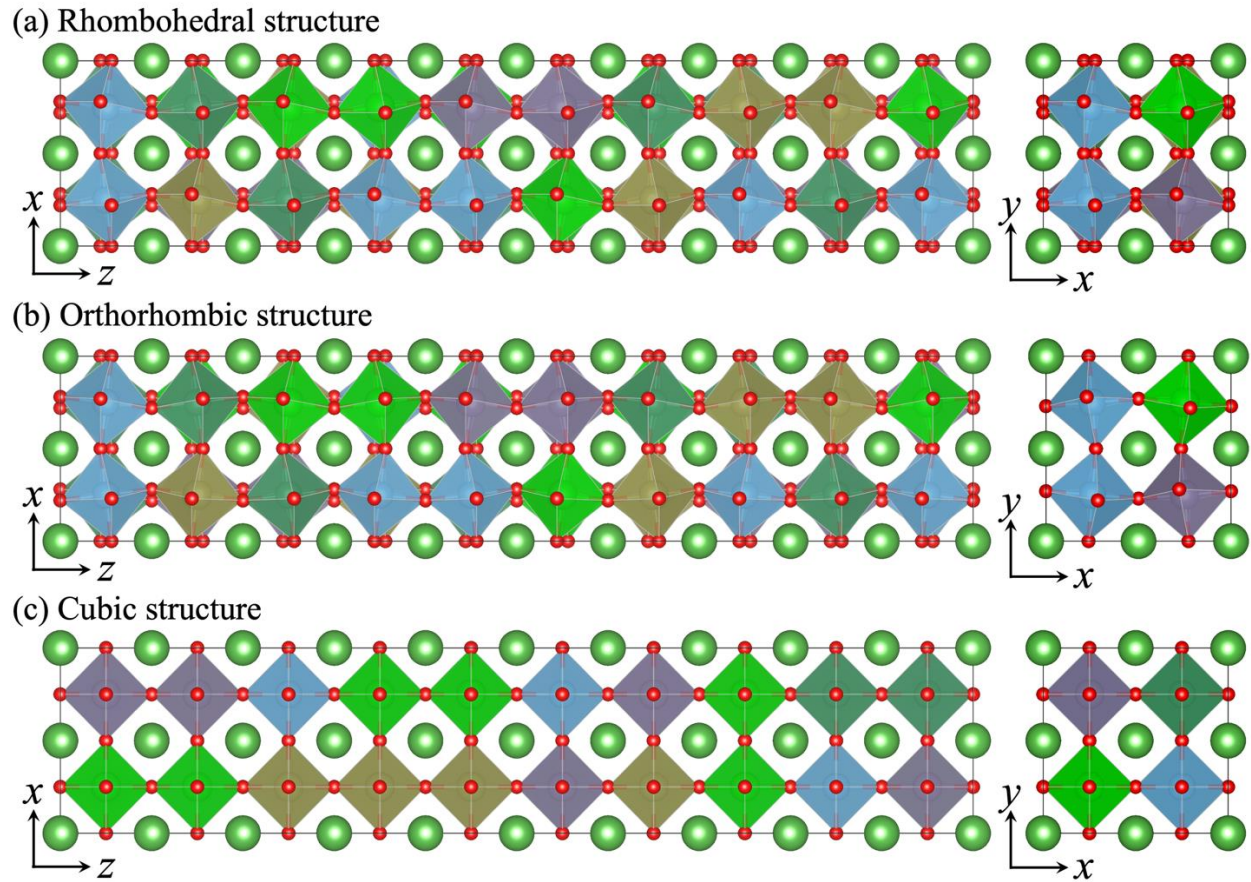

**Fig. S1**. The $2 \times 2 \times 10$ SQS models, showing the octahedral rotation patterns and $B-$site disorder, in L5$B$O for rhombohedral (a), orthorhombic (b) and cubic (c) structures. The color of the octahedra represent variation in $B-$site cations. La is the larger solid green colored sphere and oxygen is the smaller red colored sphere. The octahedral rotational patterns for all structures are identical when viewed along the *x* and *z* directions.

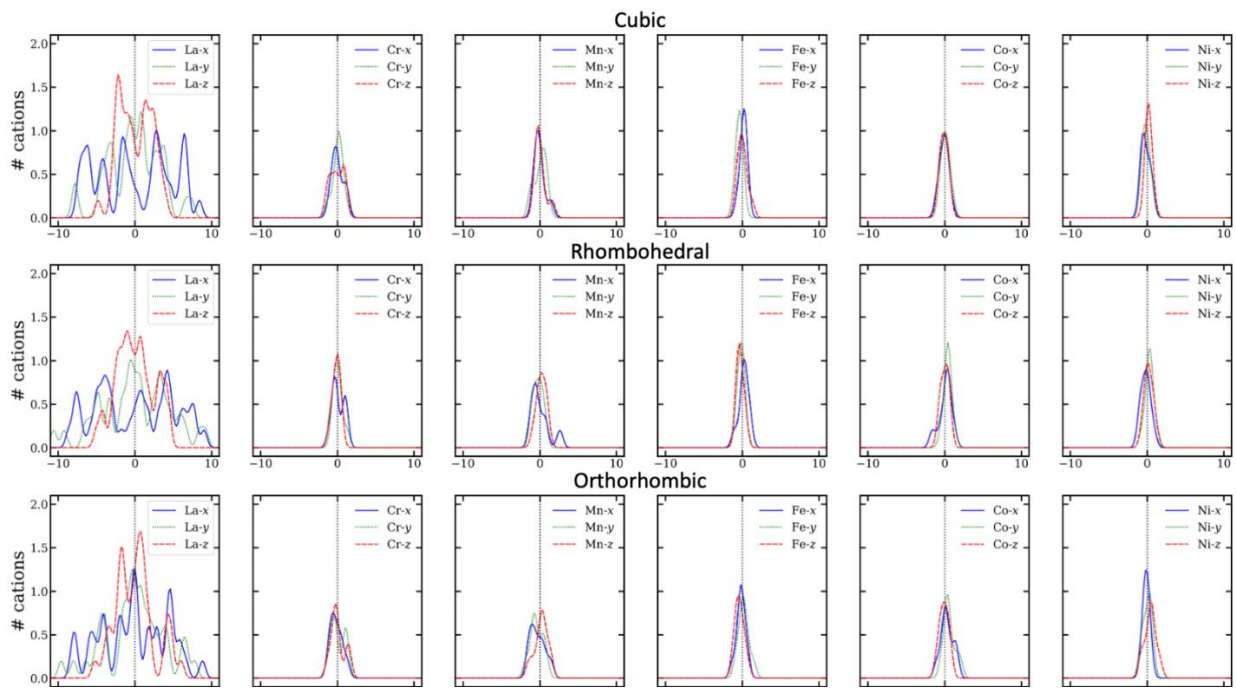

**Fig. S2**. The distribution of polar displacements for the *A*-site and *B*-site cations for the cubic (top), rhombohedral (middle) and orthorhombic (bottom) structures.

**Methods: Experimental**

The 47 nm L5*B*O film used in this work was grown using pulsed laser epitaxy (PLE) on a 5×5 mm$^2$ SrTiO$_3$ (001), (La$_{0.3}$Sr$_{0.7}$)(Al$_{0.65}$Ta$_{0.35}$)O$_3$ (LSAT) and NdGaO$_3$ (110). The stoichiometric ceramic target was synthesized using solid state reaction methods as detailed in Ref. [12]. Prior to growth, the substrate was treated using the standard recipe to achieve an atomically flat surface by etching in hydrofluoric acid followed by an anneal in air to 1050 °C for ~2 hours. In the PLE chamber, the substrate was then heated to 625 °C in 90 mTorr of oxygen where the film was deposited using a 248 nm excimer laser operating at a 5 Hz pulse repetition rate. The laser fluence was approximately 1 Jcm$^{-2}$ and the distance between the substrate and the target was about 5 cm. After growth, the sample was cooled in 200 Torr oxygen. The sample thickness was measured using X-ray diffraction and reflectivity with an estimated error less than ±1 nm. X-ray measurements were performed using a Malvern Panalytical X'Pert³ using a 4-circle goniometer. The incident beam was Cu radiation and the $k_{\alpha 1}$ line was isolated using a double bounce Ge111 monochromator and focused to a rectangular beam with width of 5 mm (orthogonal to the scattering plane) and a 0.3 mm height (parallel to the scattering plane). A linear PIXcel 2D detector was used. Intensity maps of the half order and off-axis peaks were obtained by measuring coupled $2\theta$ and $\omega$ scans with collection times between 100-250 seconds per point; within the same scan, a region adjacent to the peak was used to subtract the background, which, together, enable reliable quantification of very weak half-order reflections. A beam footprint correction was calculated using a numerical integration of the projection of the beam onto the sample, as detailed next.

## Beamfoot print correction

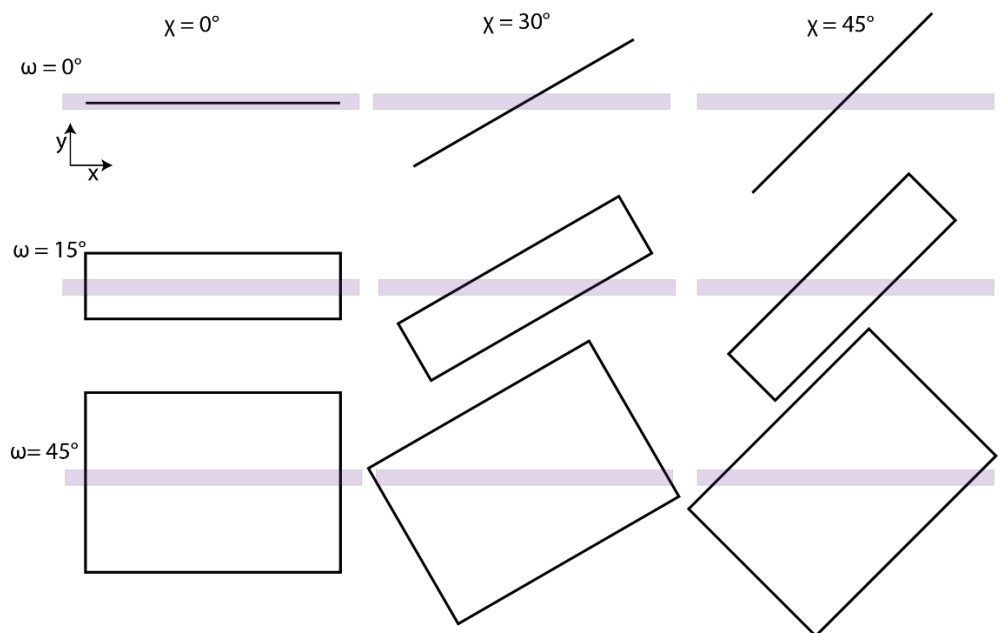

**Fig. S3** Beamfoot print correction schematic where the overlap of the beam on the sample was calculated numerically. This is illustrated looking down the beam for various values of $\omega$ (rotation axis orthogonal to the beam and parallel to the $2\theta$ axis) and $\chi$ (rotation axis parallel to the beam). The light purple rectangle is the beam and the white square is the sample.

## Results for simulated fitting

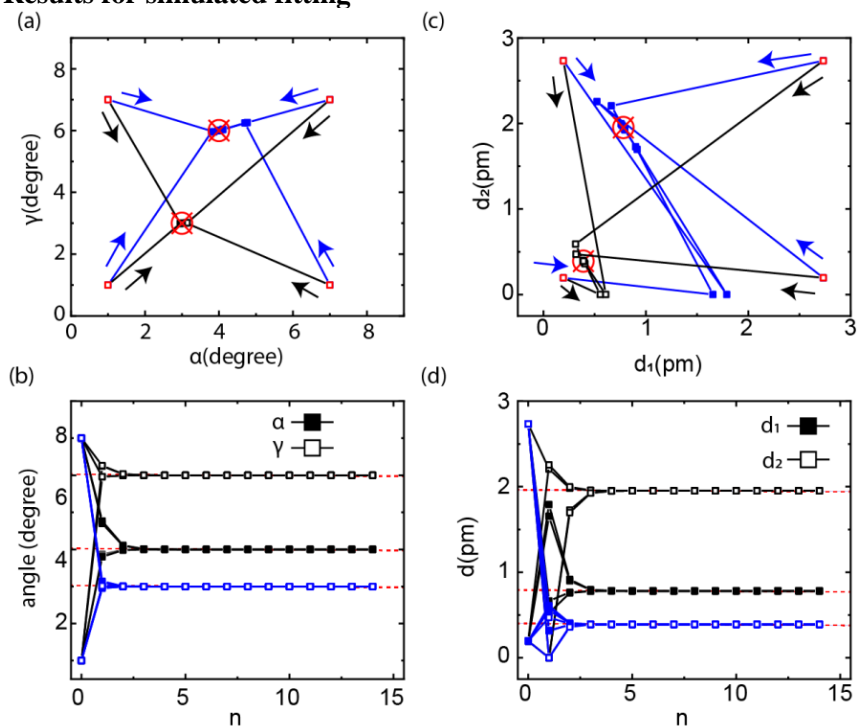

**Fig. S4** Results for testing the fitting algorithm by fitting a set of calculated intensities using the same set of reflections shown in Fig. 5(a). (a-b) $\gamma$ versus $\alpha$ (a), and $\gamma$ and $\alpha$ versus $n$ (c), where $n$ is the iteration

number during the fitting algorithm. (c-d) $d_1$ versus $d_2$ (c) and $d_1$ and $d_2$ versus $n$ (d). In (a,c) the red squares are the starting values for the fit algorithm and the red crossed-circle is the value used to generate the calculated intensities. In (b,d) the red dashed lines are the values used to generate the calculated intensities, corresponding to the red crossed-circles in (a,c).

**Debye-Waller factor**

It was found that peaks with large $q$ were not adequately fit unless a Debye-Waller factor was included. This accounts for the thermal motion of the cations that reduces intensity through diffuse scattering. We used the following model for the prefactor in equation (3) in the main text (following Ref. [13]).

$$M_j = e^{-B_j q^2}, \quad (S1)$$

where $j$ signifies the atom ($j$ = La (for the $A$-site), Cr, Mn, Fe, Co, Ni (for the $B$-site) or O) in the unit cell. $B_j$ is given by

$$B_j = \frac{11492T}{A_{m,j}\Theta^2}\phi(x) + \frac{2873}{A_{m,j}\Theta},$$

Where $A_{m,j}$ is the atomic mass, $T$ is the temperature measured in Kelvin, $\Theta$ is the Debye temperature in Kelvin, and $\phi(x)$ is given by the following integral

$$\phi(x) = \frac{1}{x}\int_0^x \frac{y}{e^y - 1} dy \quad (S2)$$

where $x = \Theta/T$. $\Theta$ was determined to be around 400 K based on fitting the 1/2 1/2 7/2, 1/2 1/2 9/2, and 3/2 3/2 3/2 peaks, which appeared most sensitive.